\documentclass{iau}

\usepackage{amsmath}
\usepackage{graphicx}
\usepackage{multirow}

\begin{document}

\lefttitle{GSE Star Formation History}
\righttitle{Gaia-Sausage-Enceladus star formation history as revealed by detailed elemental abundances}

\jnlPage{1}{7}
\jnlDoiYr{2025}
\doival{10.1017/xxxxx}
\volno{395}
\pubYr{2025}
\journaltitle{Stellar populations in the Milky Way and beyond}

\aopheadtitle{Proceedings of the IAU Symposium}
\editors{J. Mel\'endez,  C. Chiappini, R. Schiavon \& M. Trevisan, eds.}

\title{Gaia-Sausage-Enceladus star formation history as revealed by detailed elemental abundances}

\author{H. Ernandes$^{1}$, D. Feuillet$^{1,2}$, S. Feltzing$^{1}$, and \'A. Sk\'ulad\'ottir$^{3}$}

\affiliation{1. Lund Observatory, Department of Geology, Lund University, S\"olvegatan 12, Lund, Sweden \\
   2. Observational Astrophysics, Department of Physics and Astronomy, Uppsala University, Uppsala, Sweden \\
   3. Dipartimento di Fisica e Astronomia, Universitá degli Studi di Firenze, Sesto Fiorentino, Italy}

\begin{abstract}
The Gaia-Sausage-Enceladus was the last major merger and central turning point in the Milky Way's story. This event, comparable in mass to the Large Magellanic Cloud today, left behind significant debris that provides valuable insights into the assembly history of our Galaxy and the chemical evolution of dwarf galaxies.
By examining the aftermath of the GSE merger, we can delve deeper into understanding how the Milky Way's formation unfolded and how dwarf galaxies evolved chemically. Specifically, the distinct patterns of neutron capture elements such as Eu and Ba, along with Mg, offer clues about the star formation history.
Through a comprehensive analysis of data compiled in the SAGA database, we investigated the Gaia Sausage-Enceladus' star formation history. Elemental abundance ratios ([Eu/Mg], [Ba/Mg], and [Eu/Ba]) derived from this study, when compared with those of surviving Milky Way satellites, indicate that the GSE experienced a prolonged period of slow star formation, lasting over 2 Gyr, until it was eventually quenched by merging with the Milky Way.
Consequently, these elemental signatures serve as a unique window into the complex history of both surviving and accreted satellites orbiting our Galaxy.
\end{abstract}


\maketitle

\section{Introduction}

%Our page limits are flexible: 6 to 14 pages for invited reviews, 5 to 10 for invited talks and 4 to 8 for contributed talks. Let us know if you need some more space. Posters should be 2 pages (but 3 is OK). Your contribution must be prepared using the file iauguide_iau395.tex, included in the attached zip directory, which contains all relevant files.
%The deadline is January 22nd, but if you are on vacation, plan to send your contribution in advance, for example, before Christmas.

Beatriz Barbuy and I met when I was a young student at the University of São Paulo during a meeting about neutrinos in astrophysics. This meeting was recommended by my friend Fernanda Hüller, knowing my interests in nuclear physics, particles, and astrophysics.
Meeting Beatriz that day played a central role in my career. She invited me to undertake a scientific initiation project with her, guiding me through the complexities of stellar astrophysics and unveiling the beauty of astronomy. Her mentorship redirected my focus from particle physics to astrophysics.
Through her mentorship, I learned that scientific work is not just about results and hard work but also about collaboration. Beatriz's enthusiasm for the field and her dedication to precise and meticulous research have been crucial in shaping my approach to research and collaboration. Working with Beatriz, I explored the intricate processes governing stellar populations, particularly nucleosynthesis from light to heavy elements, a passion I continue to pursue and now apply to the Gaia-Sausage-Enceladus research. 

Recently, using astrometric data from ESA’s Gaia satellite, researchers confirmed the presence of the Gaia-Sausage-Enceladus, which merged with the Milky Way about 10 billion years ago (Helmi et al. 2018; Belokurov et al. 2018). Its distinct chemical composition reflects its unique star formation history. $\alpha$-elements like oxygen and magnesium, produced in core-collapse supernovae on short timescales, differ from iron, which is mainly generated by type Ia supernovae over longer times. Heavy elements such as barium and europium, created through neutron-capture processes, further reveal the galaxy's complex enrichment history.

By examining elemental abundances in stars from Gaia-Sausage-Enceladus, we can trace its star formation history as seen by the elemental abundances. This work builds on studies like Sk\'ulad\'ottir and Salvadori (2020), combining nucleosynthesis timescales with observational data to understand the early stages of this ancient galaxy. Using a compilation of literature data, including Mg, Ba, and Eu abundances across a wide metallicity range, we constrain the galaxy's evolution and its contributions to the Milky Way.

\section{Data}

\subsection{Chemical Abundances in SAGA}

\begin{figure*}
    \centering
    \includegraphics[width=4in]{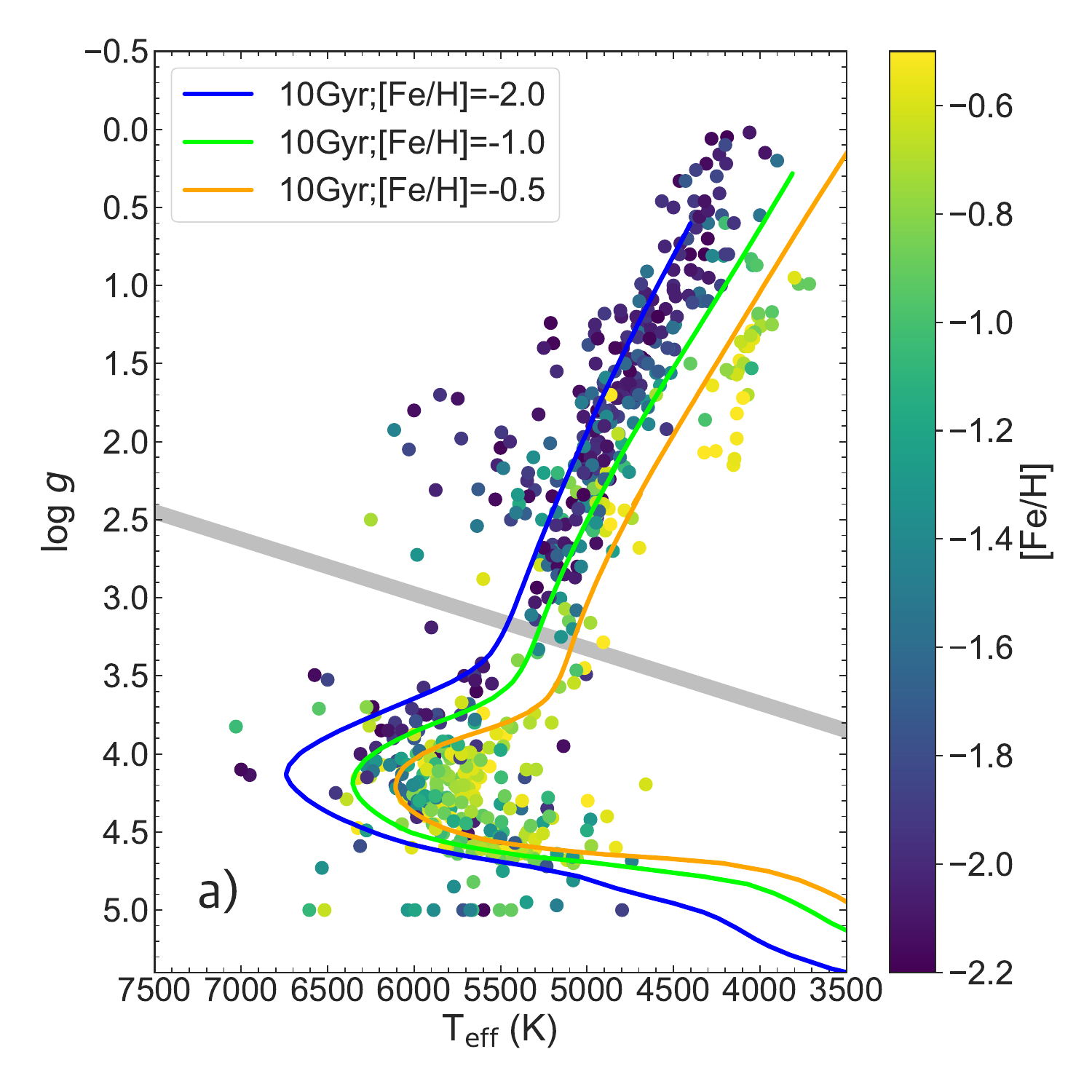}
    \caption{Properties of the stars in our final sample with SAGA and Gaia data, with three 10 Gyr PARSEC isochrones overlaid, metallicity as indicated in the legends: Kiel diagram, colour-coded by the median [Fe/H]. The shaded grey line represents the visual definition we used in fig. \ref{Diane-trends} to divided the sample into dwarf and giant stars.}
    \label{fullCMDHR}
\end{figure*}

This study aims to reconstruct the star formation history of Gaia-Sausage-Enceladus using elemental abundance trends. Specifically, we focus on neutron-capture elements and their trends at low metallicity, reflecting early star formation stages. Previous work by Matsuno et al. (2021) analyzed $r$- and $s$-process elements in Gaia-Sausage-Enceladus for $\rm -1.7<[Fe/H]<-0.4$. In order to reach the metal-poor end, that provides a valuable resource for studying the abundance trends we use the SAGA database (Suda et al. 2008) which compiles elemental abundances from literature sources indexed in the CDS\footnote{Centre de Données astronomiques de Strasbourg} database. 

Through the SAGA plotting interface\footnote{http://sagadatabase.jp/}, we selected stars meeting the following criteria: \\
\begin{itemize} \item[i)] $-2.2 <$ [Fe/H] $< -0.5$,
\item[ii)] Measurements available for both [Eu/Fe] and [Fe/H]. \\
\end{itemize} 

This selection yielded 804 stars. After identifying \textit{Gaia} DR3 IDs, some duplicates were resolved (Sect.\ref{sec:Gaia}). For these stars, we supplemented [Mg/Fe] and [Ba/Fe] measurements when available and excluded Ba-rich stars, as they result from binary transfer rather than tracing chemical evolution.

The final catalogue assigns a unique row for each literature source reporting [Fe/H], [Eu/Fe], [Mg/Fe], and [Ba/Fe], allowing multiple entries per star. Elemental abundances and stellar parameters ($T_{\rm eff}$, log$g$, [Fe/H]) may vary across studies. To address multiple abundance values for a single element in the same star, we adopted the median value. 

Figure~\ref{fullCMDHR} presents the Kiel diagram for the final sample of 654 stars with SAGA and \textit{Gaia} parameters. Median $T_{\rm eff}$, log$g$, and [Fe/H] values were used. Metal-rich giant branch stars, predominantly from Tautvaisiene et al. (2021), are separated due to their higher metallicities ($\rm [Fe/H]>-0.7$).

\subsection{Astrometric Data and Kinematics} 
\label{sec:Gaia}

We utilize \textit{Gaia} astrometric parameters to compute the kinematics of the stars in our sample. First, we retrieve \textit{Gaia} DR3 IDs using the \textsc{astroquery.simbad} package (Ginsburg et al. 2019; Wenger et al. 2000) in Python. During this process, we identify 13 stars from Li et al. (2022), included in our SAGA sample, that are not listed in the SIMBAD database under the IDs provided by SAGA. After consulting Haining Li, we obtained the \textit{Gaia} DR3 IDs for these stars. This results in 680 unique stars from SAGA with \textit{Gaia} DR3 IDs.

Subsequently, we query the \textit{Gaia} archive\footnote{https://gea.esac.esa.int/archive/} for \textit{Gaia} DR3 parameters (Gaia Collaboration and Brown 2021) and find that 654 stars have available astrometric and Radial Velocity Spectrometer (RVS) data. Using the Python packages \textsc{astropy} (Astropy Collaboration et al. 2013, 2018, 2022) and \textsc{galpy} (Bovy 2015), we compute the full kinematics and orbital parameters for these stars. For action calculations, we employ the \textsc{actionAngleStaeckel} approximation (Binney 2012; Bovy and Rix 2013) with a delta value of 0.4. We perform these calculations using both the \textsc{MWPotential14} (Bovy and Rix 2013) and McMillan (2017) Milky Way potential models to facilitate comparisons with various Gaia-Sausage-Enceladus selection methods reported in the literature. Geometric distances from  Bailer-Jones et al. (2021) are adopted.

For spectroscopic radial velocities, we prioritize measurements from Gaia RVS. If not available, we use SAGA radial velocities, where there are multiple radial velocity entries for the same existing star. We adopt the most recent value, as deviations between literature values are minimal. 

\section{Analysis}

\begin{figure*}
    \centering
    \includegraphics[width=0.95\textwidth]{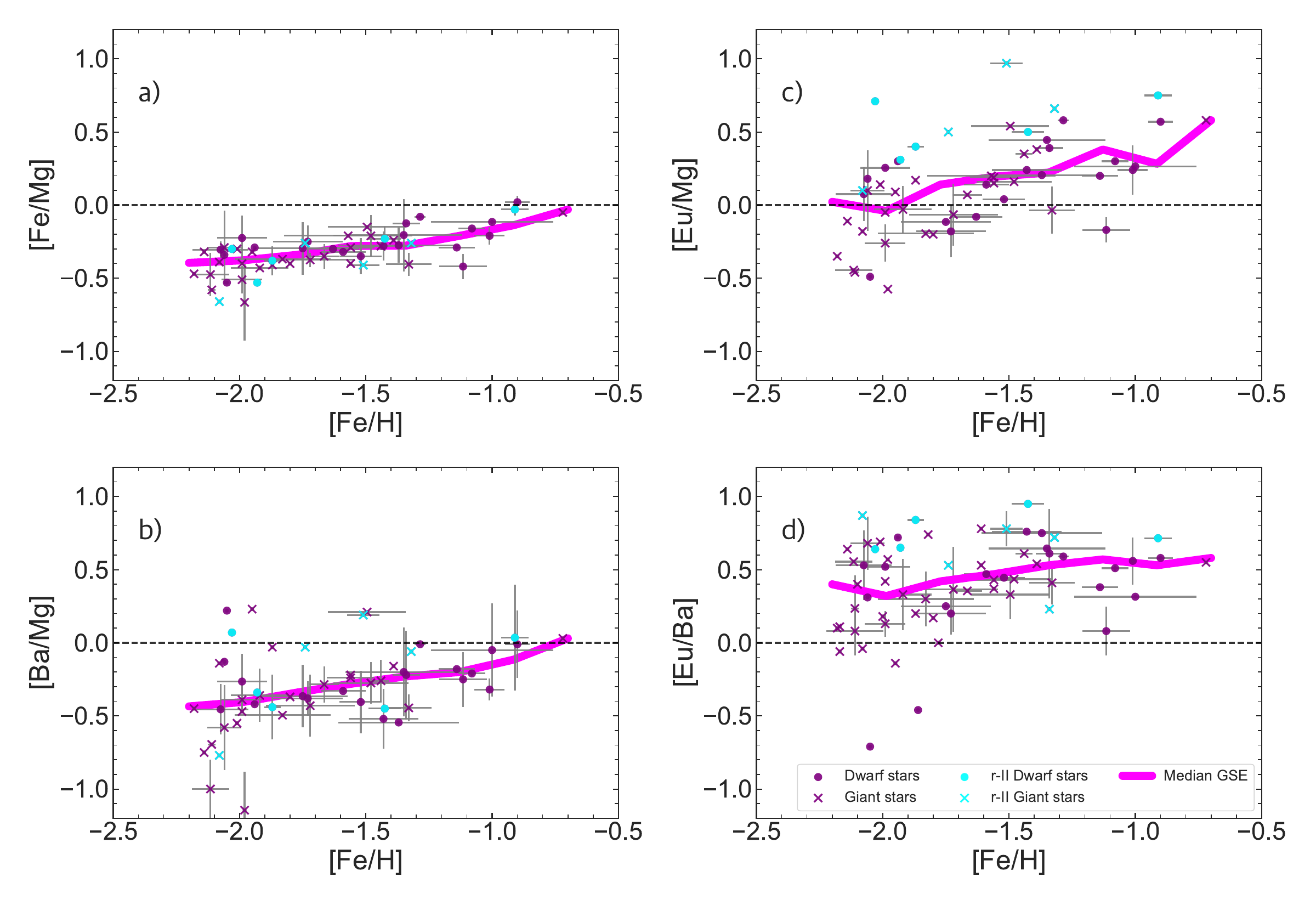}
    \caption{
        Elemental abundance ratios as a function of [Fe/H] for Gaia-Sausage-Enceladus stars selected from the SAGA database. Panels display: a) [Fe/Mg]; b) [Ba/Mg]; c) [Eu/Mg]; and d) [Eu/Ba]. Symbols indicate dwarfs (circles) and giants (x-symbols), with $r$-II stars shown in cyan and $r$-normal stars in purple. Pink lines represent the median trends. 
    }
    \label{Diane-trends}
\end{figure*}

Several methods for identifying Gaia-Sausage-Enceladus members have been proposed in the literature (e.g., Belokurov et al. 2018; Helmi et al. 2018; Myeong et al. 2019; Feuillet et al. 2021; Horta and Schiavon 2023). In Fig.\,\ref{Diane-trends} we show our primary selection method is the scheme from Feuillet et al. (2021), chosen for its demonstrated balance of purity and inclusivity in identifying Gaia-Sausage-Enceladus members according to (Carrillo et al. 2024).. Using this scheme, we identify 73 stars with both Fe and Eu measurements from our SAGA sample. Of these, 70 have [Ba/Fe] measurements, 61 have [Mg/Fe] measurements, and 59 have measurements for both [Ba/Fe] and [Mg/Fe]. 

\section{Results: Elemental Abundance Trends}
\label{sect:result}

\begin{figure*}
    \centering
    \includegraphics[width=0.95\textwidth]{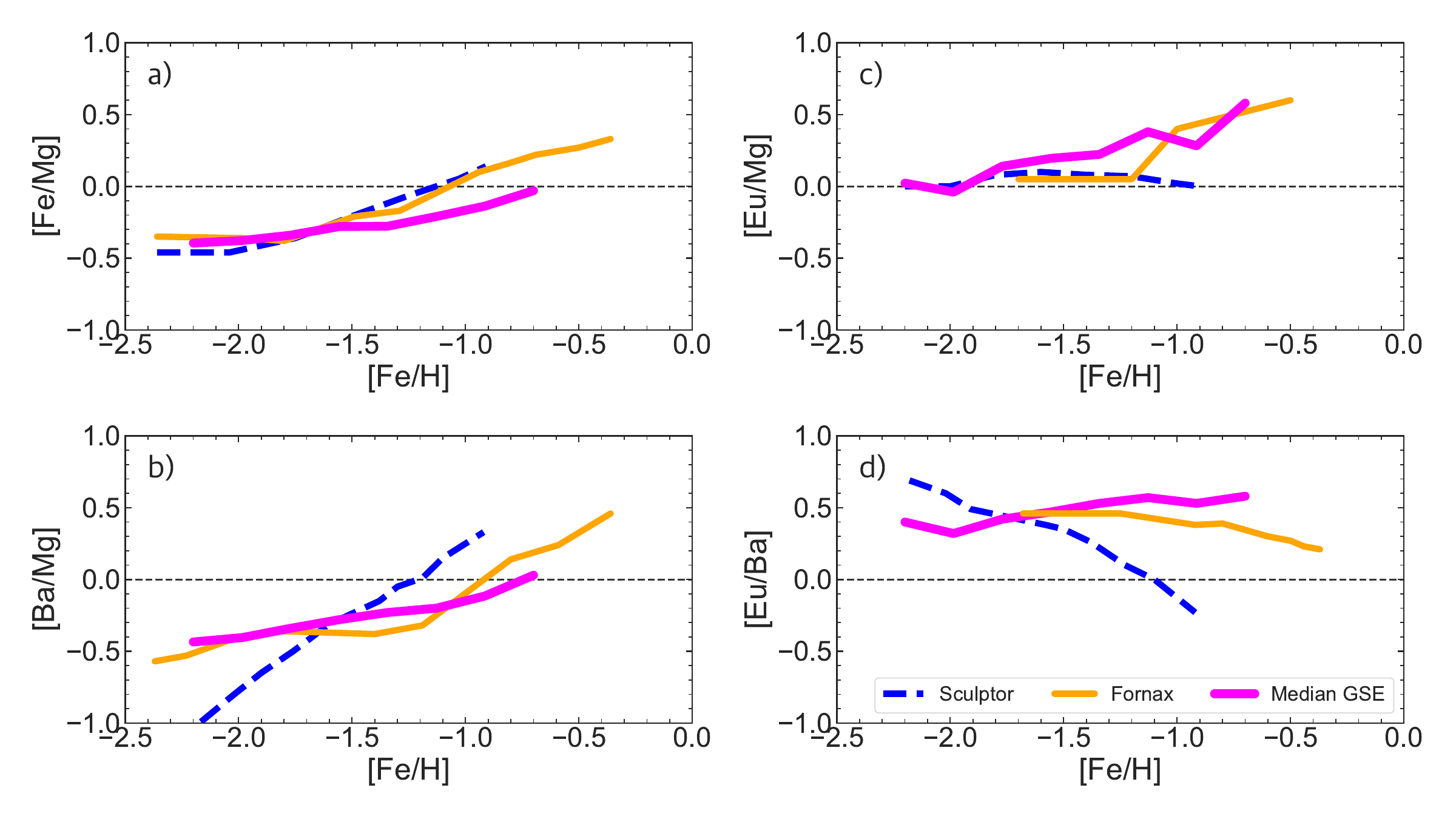}
    \caption{
    Median abundance trends of stars in the Gaia-Sausage-Enceladus (Fig.~\ref{Diane-trends}) in pink solid line, and the dSph galaxies Fornax (orange solid line) and Sculptor (blue dashed line) from Sk\'ulad\'ottir and Salvadori (2020). Abundance ratios are shown as a function of [Fe/H] for: a) [Fe/Mg]; b) [Ba/Mg]; c) [Eu/Mg]; and d) [Eu/Ba].
  }
    \label{Dwarfs-trends}
\end{figure*}

Figure~\ref{Diane-trends} presents four elemental abundance ratios ([Fe/Mg], [Ba/Mg], [Eu/Mg], and [Eu/Ba]) for 73 stars selected following Feuillet et al. (2021).  These include both dwarfs and giants and highlight $r$-II stars. The abundance trends for dwarfs and giants are consistent, with $r$-II stars generally following the median trends except for their enhanced Eu abundances. Median trends are calculated using a running median over 0.25\,dex [Fe/H] bins with 0.15\,dex steps, excluding error bars.

Across all abundance ratios ([Fe/Mg], [Ba/Mg], [Eu/Mg], [Eu/Ba]), trends increase with [Fe/H]. However, [Eu/Mg] and [Eu/Ba] show more complex, non-monotonic behaviour. [Fe/Mg] exhibits the least scatter around its median trend, while abundance ratios involving Eu display larger star-to-star scatter. This scatter may partly arise from systematic differences between studies, as most stars in the SAGA database have only one Eu measurement available. A detailed, homogeneous high-resolution spectroscopic analysis across the metallicity range is necessary to assess the intrinsic scatter robustly. In contrast, [Ba/Mg] trends show less scatter, with larger deviations at lower [Fe/H].

Stars selected as Gaia-Sausage-Enceladus members using  Myeong et al. (2019) exhibit trends similar to those of Feuillet et al. (2021), with higher Eu abundances at $\rm[Fe/H] \gtrsim -1.5$. This suggests fewer Milky Way stars in the Myeong et al. (2019)  sample. However, since the Feuillet et al. (2021) selection includes twice as many stars and produces comparable trends, it is adopted for further analysis. Carrillo et al. (2024) demonstrated its higher purity using cosmological simulations of Milky Way-like galaxies.

\section{Discussion}

Our primary goal is to characterize the star formation history of the Gaia-Sausage-Enceladus progenitor galaxy using elemental abundance trends from its stellar debris. Assuming the debris originated from a single, ancient, low-mass galaxy, we interpret these trends as representative of a distinct star formation history and chemical enrichment process. By analyzing observed abundance ratios ([Fe/Mg], [Ba/Mg], [Eu/Mg], and [Eu/Ba]) and the enrichment timescales of Mg, Fe, Ba, and Eu, we compare Gaia-Sausage-Enceladus to the Sculptor and Fornax dwarf galaxies (Fig.~\ref{Dwarfs-trends}), which provide contrasting star formation histories: Sculptor experienced a brief, intense burst, while Fornax had a prolonged, gradual history.

\subsection{Enrichment Timescales of Mg, Fe, Ba, and Eu}

Mg is synthesized in core-collapse supernovae (ccSN) and reflects rapid star formation (Woosley and Weaver 1995; Woosley and Heger 2007). Fe originates from both ccSN and delayed Type Ia supernovae (SN Ia), which enrich on timescales of 0.1--2 Gyr  (Maoz et al. 2014; Palicio et al. 2024). Ba is predominantly produced in the $s$-process within Asymptotic Giant Branch (AGB) stars, reflecting delayed and extended enrichment (Karakas and Lattanzio 2014). Eu, formed via the $r$-process, has both quick and delayed sources, such as neutron star mergers (Abbott et al. 2017; Sk\'ulad\'ottir et al. 2019).

\begin{itemize} 

\item{[Fe/Mg] and Star Formation Truncation:}
The [Fe/Mg] ratio traces the balance between rapid ccSN enrichment and delayed SN Ia enrichment. Gaia-Sausage-Enceladus does not show positive [Fe/Mg] at high [Fe/H], unlike Fornax and Sculptor, where a gradual decline in star formation is evident. This absence suggests star formation in Gaia-Sausage-Enceladus ceased abruptly rather than declining naturally due to gas depletion.

\item{[Ba/Mg] and Initial Star Formation:}
[Ba/Mg] reflects the interplay of quick ccSN enrichment and delayed AGB enrichment. At low [Fe/H], Gaia-Sausage-Enceladus trends resemble Fornax, with no strong initial burst of star formation, unlike Sculptor. The smooth increase in [Ba/Mg] with [Fe/H] indicates extended star formation, but the absence of a high [Fe/H] tail suggests a sudden quenching.

\item{[Eu/Mg] and Extended Star Formation:}
The flat [Eu/Mg] trend at low [Fe/H] in Gaia-Sausage-Enceladus implies inefficient star formation during the first $\sim 2$ Gyr, consistent with [Ba/Mg]. Above $\rm[Fe/H] \approx -2.0$, [Eu/Mg] increases smoothly, indicating contributions from both quick and delayed $r$-process sources, suggesting ongoing enrichment without dramatic changes in star formation.

\item{[Eu/Ba] and Quenching Evidence:}
The [Eu/Ba] ratio reflects the balance between delayed AGB enrichment (Ba) and $r$-process sources (Eu). At low [Fe/H], Gaia-Sausage-Enceladus shows lower [Eu/Ba] than Sculptor, consistent with weak initial star formation. Unlike Fornax, Gaia-Sausage-Enceladus exhibits a rising [Eu/Ba] at higher [Fe/H], suggesting stable star formation until its abrupt cessation.

\end{itemize}

\begin{figure*}[h]
    \centering
    \includegraphics[width=0.95\textwidth]{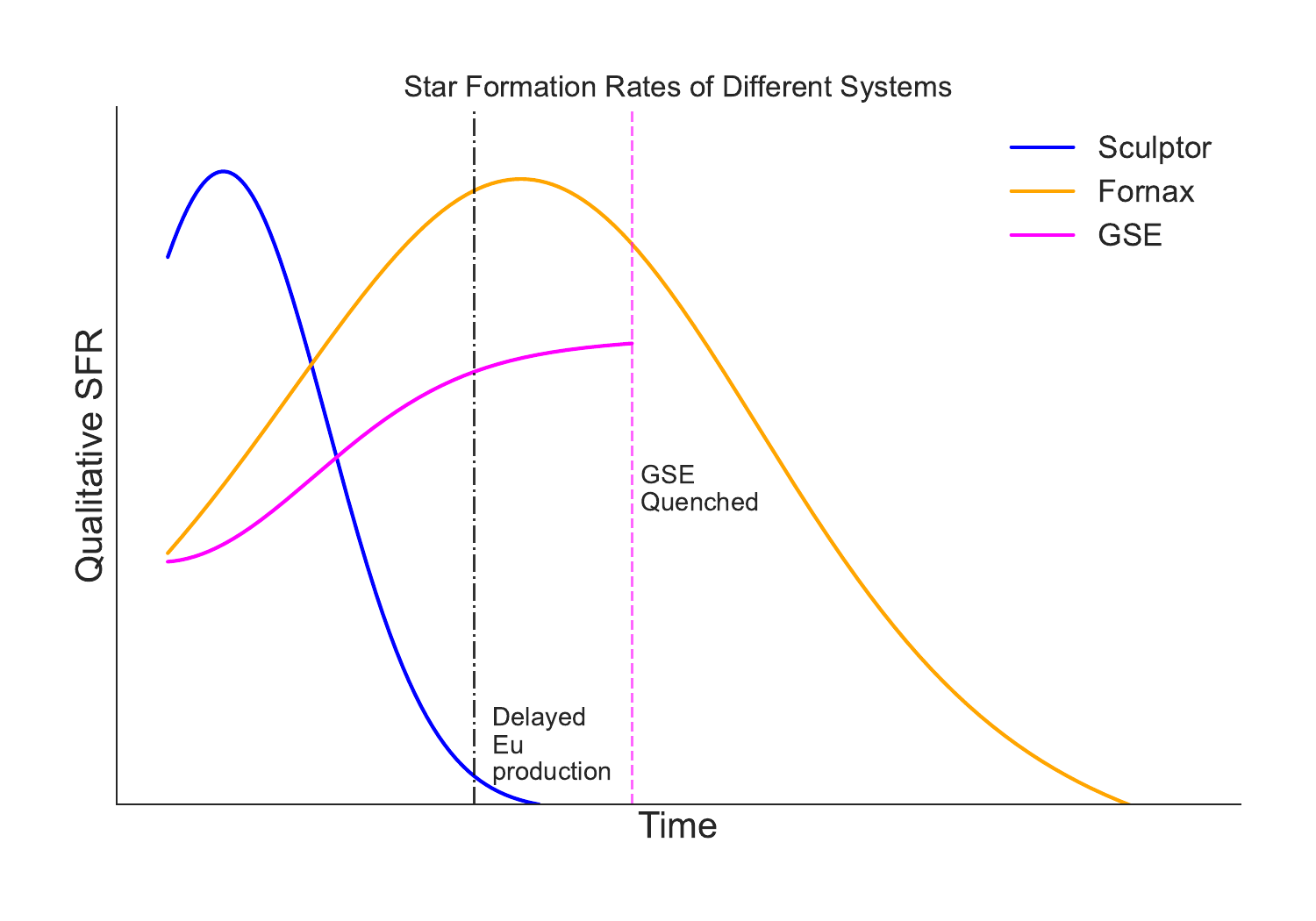}
    \caption{ Qualitative star formation rate over time for Sculptor in blue, Gaia-Sausage-Enceladus (GSE) in magenta, and Fornax in yellow. }
    \label{SFH}
\end{figure*}

Figure \ref{SFH} presents a qualitative schematic of the star formation history for Gaia-Sausage-Enceladus. This schematic is shown alongside the qualitative star formation histories of Sculptor and Fornax, based on the elemental abundance trends analysed in Sk\'ulad\'ottir and Salvadori (2020) for Sculptor and Fornax, and in Ernandes et al. (2024) for Gaia-Sausage-Enceladus. Additionally, Figure \ref{SFH} indicates the point at which the delayed source of Eu begins contributing to their star formation histories.

%%%%%%%%%%%%%%%%%%%%%%%%%%%%%%%%%%%%%%%%%%%%%%%%%%%%%%%%%%%%%%%%%%%%%%%%
\section{Conclusions}
\label{sect:conclusion}

Our study examines the elemental abundance trends of $r$- and $s$-process elements, Eu and Ba, alongside Mg and Fe, in the Gaia-Sausage-Enceladus galaxy. We analyze a sample of stars with $\rm -2.2 < [Fe/H] < -0.5$ from the SAGA database, focusing on the elemental abundance ratios [Fe/Mg], [Ba/Mg], [Eu/Mg], and [Eu/Ba] to constrain the star formation history of the progenitor galaxy. Comparing these trends with Fornax and Sculptor, we establish three key constraints on the star formation history of Gaia-Sausage-Enceladus:

1) The high [Ba/Mg] and low [Eu/Ba] at low [Fe/H] indicate that star formation was initially low, unlike Sculptor, where a strong burst led to rapid enrichment from core-collapse supernovae (ccSN) and asymptotic giant branch (AGB) stars.

2) The increase in [Eu/Mg] suggests star formation lasted beyond $\sim$2~Gyr, continuing until delayed $r$-process enrichment occurred. This contrasts with Sculptor, where star formation did not last long enough for [Eu/Mg] to increase, and Fornax, where extended star formation resulted in high [Eu/Mg] at high [Fe/H].

3) The lack of a decrease in [Eu/Ba] at high [Fe/H] in Gaia-Sausage-Enceladus implies that its star formation was quenched at $\rm [Fe/H] \sim -0.5$, preventing a transition to a regime dominated by Ba enrichment from AGB stars.

The maximum values of [Fe/Mg] and [Ba/Mg] in Gaia-Sausage-Enceladus are $\sim$0, unlike the super-solar values in Fornax and Sculptor, suggesting that Gaia-Sausage-Enceladus did not naturally exhaust its gas supply.

Looking ahead, the upcoming 4MOST data is expected to significantly enhance our understanding of the Milky Way's accreted populations, particularly the major accretion event associated with Gaia-Sausage-Enceladus. Surveys covering the halo, disk, and bulge (Surveys 1-4; Helmi and Irwin 2019; Christlieb et al. 2019; Chiappini et al. 2019; Bensby and Bergemann 2019), along with data on dwarf galaxies and their stellar streams in Survey 14 (Sk\'uad\'ottir and et al. 2023), will provide unprecedented insights into the formation and evolution of the Milky Way with greater precision than ever before.

%35

\end{document}